\documentstyle[preprint,aps,eqsecnum,tighten]{revtex}

\def\be{\nopagebreak[3]\begin{equation}}
\def\ee{\end{equation}}
\def\ba{\nopagebreak[3]\begin{eqnarray}}
\def\ea{\end{eqnarray}}

\def\bp{<\!\Psi_c|}
\def\d{\dagger}
\def\e{\exp\, }
\def\G{{\Gamma}}
\def\kp{|\Psi_c \!>}
\def\N{{\cal N}} 
\def\o{{\omega}}

\begin{document}

\title{Large quantum gravity effects:\\ 
Unexpected limitations of the classical theory}
\author{Abhay Ashtekar\thanks{Electronic address: 
ashtekar@phys.psu.edu}}
\address{Center for Gravitational Physics and Geometry\\
Physics Department, Penn State, University Park, PA 16802, USA} 

\maketitle

\begin{abstract}

3-dimensional gravity coupled to Maxwell (or Klein-Gordon) fields is
exactly soluble under the assumption of axi-symmetry. The solution is
used to probe several quantum gravity issues. In particular, it is
shown that the quantum fluctuations in the geometry are large unless
the number and frequency of photons satisfy the inequality $\N(\hbar
G\omega)^2<\!<1$. Thus, even when there is a single photon of
Planckian frequency, the quantum uncertainties in the metric are
significant.  Results hold also for a sector of the 4-dimensional
theory (consisting of Einstein Rosen gravitational waves).

\end{abstract}
\section{}

This work has several motivations. The first stems from the fact that
recently there has been considerable mathematical progress in various
approaches to quantum gravity. Hence, it is now important to devise
criteria for the physical viability of potential solutions. One
obvious demand is that the theory should admit ``a sufficient number
of'' semi-classical states. However, the precise meaning of this
requirement is not obvious.  What exactly is the set of classical
solutions that should admit quantum analogs? Indeed, even for the
Hydrogen atom problem the answer is subtle from the perspective
of the classical theory.  In the present case, we do not have direct
experimental data to guide us.  Therefore, a natural strategy is to
try to develop intuition by analyzing exactly soluble models. Such
models can also provide insights in to a number of other issues. For
example, since the modes that are significant at infinity in the
Hawking process have trans-Planckian frequencies near the horizon,
there has been considerable interest in the role of high frequency
fields in quantum gravity. There have been suggestions that an
adequate description of such modes may violate local Lorentz
invariance \cite{1}. It has also been suggested that the usual
counting of states at high frequencies is inadequate \cite{2} and that
the correct counting may lead to a ``holographic picture'' in which
physics within a region can be captured by states residing at its
boundary \cite{3}. Finally, there exists a mathematically well-defined
theory called semi-classical gravity in which the gravitational field
is treated classically but the matter fields are quantized
\cite{4}. One hopes that this theory would be phenomenologically
satisfactory in the low energy-regime. Is this indeed the case? Can
one derive this theory from full quantum gravity?  What does it ``miss
out''?

The purpose of this letter is to analyze such issues in the context of
an exactly soluble model: 3-dimensional (Lorentzian) gravity coupled
to Maxwell fields, restricted by the condition that there be a
rotational symmetry. As is well-known, in 3 dimensions, a Maxwell
field $F_{ab}$ is dual to a scalar field $\Phi$ via $F_{ab} =
\epsilon_{ab}{}^c \nabla_c\Phi$ and Maxwell equations reduce to the
wave equation, $g^{ab}\nabla_a \nabla_b \Phi = 0$.  Furthermore, the
duality mapping sends the stress-energy tensor $T_{ab}(F)$ to
$T_{ab}(\Phi)$. Therefore, mathematically, the Einstein-Maxwell system
(without further fields and interactions) is equivalent to the
Einstein-Klein-Gordon system. For simplicity, we will use the scalar
field formulation. Since this system is exactly soluble both
classically (see e.g. \cite{5}) and quantum mechanically \cite{6,7}
under the assumption of axi-symmetry, we can explicitly analyze the
status of various quantum issues in the solution. We will find that
quantum effects can be unexpectedly large so that a very large number
of solutions of the classical theory are spurious: they do not appear
in the classical limit of the quantum theory. The model is completely
equivalent \cite{5} to a ``midi-superspace'' consisting of
Einstein-Rosen waves in {\it 4-dimensional} source-free general
relativity. Hence, our results apply also to this sector of
4-dimensional quantum gravity \cite{8}. Finally, it should not be
difficult to extend the discussion to the treatment of CGHS dilatonic
black holes given in \cite{mv}.

In order to compare and contrast various effects and better understand
their origin, let us proceed in steps, ``switching $\hbar$ and $G$ on
and off'' as needed. (However, throughout we set $c=1$.) Let us begin
with quantum field theory without gravity ($G=0$). Then we just have
an axi-symmetric quantum Maxwell field in 3-dimensional Minkowski
space-time $(M, \eta_{ab})$, represented by the
operator-valued distribution
\be \label{1}
\hat\Phi (R,T) = \int_0^\infty d\o [f_{\o}^{+}(R,T) A(\o) 
+ f_{\o}^{-} (R,T) A^\dagger (\o)]\, .
\ee
Here $f_\o^+ (R,T) = J_o(\o,R) \e{-i\o T}$ (with $J_o$, the Bessel
function) are the positive frequency solutions, $A^\dagger$ and $A$,
the creation and annihilation operators, satisfying the commutation
relations $[A(\o),\, A^\dagger(\o')] = \hbar \delta(\o, \o')$. In the
rest frame used in the expansion (\ref{1}), the Hamiltonian is given
by:
\be \label{2}
\hat{H}_o = \int_0^\infty d\o\, \o A^\d(\o) A(\o)\, . 
\ee
The system is Poincar\'e invariant and other Poncar\'e generators can
be expressed similarly. The Hilbert space is the Fock space. Given
{\it any} classical solution,
\be \label{3}
\tilde{C} (R,T) = \int_0^\infty d\o [f_{\o}^{+}(R,T) C(\o) 
+ f_{\o}^{-} (R,T) \bar{C} (\o)],
\ee
there is a (normalized) coherent state, $\kp$ which remains peaked
around $\tilde{C}$ for all times:
\be \label{4}
\kp \,=\, \big(\e -\frac{1}{2} \int_0^\infty d\o\, |C(\o)|^2 \big)\,\,
\e \frac{1}{\hbar} \int_0^\infty d\o\, C(\o) A^\d (\o)\,\, |0\!>
\ee
However, quantum fluctuations in physical quantities are often
negligible compared to the expectation value of that quantity only
when the the expected number of photons, 
\be \label{5}
\N \,=\, \bp \hat{N} \kp \,\,\equiv \,\,\frac{1}{\hbar}\int d\o 
|C(\o )|^2 \, ,
\ee
is large. For example, we have: $(\triangle \hat{H}_o/\!<\!\hat{H}_o
\!>)^2 \sim 1/\N$, where, as usual, $(\triangle \hat{H}_o)^2 =
<\hat{H}_o^2> - <\hat{H}_o>^2$ is the uncertainty in the value of the
Hamiltonian.  The entire discussion is independent of where $C(\o)$
may be peaked; since there is no preferred scale, the theory can not
distinguish high frequency photons from the low frequency ones.

Let us now switch off $\hbar$ and switch on $G$; i.e. consider
classical general relativity. (Since in 3-dimensions, $G$ has the
physical dimensions of inverse mass, we now have a mass scale which
features prominently in the description.) The theory is now governed
by a set of {\it coupled, non-linear} partial differential equations
on $M \equiv R^3$:
\be \label{6}
g^{ab} \nabla_a \nabla_b \Phi = 0 \quad {\rm and}\quad R_{ab} = 
8\pi G \,\, \nabla_a\Phi \nabla_b\Phi\, ,
\ee
where $R_{ab}$ is the Ricci tensor of $g_{ab}$. 

However, because of the rotational invariance {\it and} presence of
the axis, two simplifications arise \cite{bbetal,5}. First, $M$
admits a global chart $T, R, \theta$ such that the norm of the
rotational Killing field $\partial/\partial \theta$ is $R^2$. (The
chart is unique up to translations $T \rightarrow T +k$, where $k$ is
a constant.)  In this chart, the metric has the form:
\be \label{7}
g_{ab}dx^a dx^b = e^{G\G (R,T)}(-dT^2 + dR^2) + R^2 d\theta^2\, ;
\ee
there is only one unknown metric coefficient, $\G (R,T)$. The second
simplification is that $\Phi$ satisfies the wave equation with respect
to $g_{ab}$ if and only if it satisfies the wave equation with respect
to the flat metric $\eta_{ab}$ obtained by setting $\G (R, T)=0$ in
(\ref{7}). Thus, the two equations in (\ref{6}) can now be {\it
decoupled}: we can first solve for $\Phi$ and then attempt to solve
the Einstein's equation to determine $\G(R,T)$.  Given a solution
$\tilde{C}(R,T)$ to the wave equation, the corresponding $\G (R,T)$
turns out to be:
\be \label{8}
\G (R,T) = \frac{1}{2} \int_0^R dR R\, \big[(\partial_T\, 
\tilde{C}(R,T))^2 + (\partial_R \, \tilde{C}(R,T))^2 \big]\, ,
\ee
which is precisely the (Minkowskian) energy of the scalar field
$\tilde{C}(R,T)$ in a box of radius $R$ at time $T$. Thus, the problem
of solving the coupled Einstein-Maxwell system reduces to that of
solving the wave equation on the flat space $(M, \eta_{ab})$.  In the
Hamiltonian language, one can fully gauge fix the system, show that the
true degree of freedom can be made to reside in the scalar field,
focus on its dynamics and treat the metric as a secondary quantity
which can be expressed, at the end, as a functional of the scalar
field \cite{7}. As in the previous case (with $\hbar\not= 0,\, G =0$),
there is no natural frequency scale and the general description in
insensitive to the detailed profile of $C(\o)$. The asymptotic metric,
for example, depends only on the total (Minkowskian) energy
$H_0(\tilde{C})$ and not to where $C(\o )$ is peaked.

Suppose the field $\tilde{C}(R,T)$ has initial data of compact
support. Then, a neighborhood of infinity is source-free and {\it in
that region} the metric is locally flat, with $\G (R,T) =
H_o(\tilde{C})$, a constant (given by the total energy of
$\tilde{C}$). Nonetheless, when $GH_o(\tilde{C}) \ge 1$, the
asymptotic geometry is {\it quite different} from that of the globally
flat metric $\eta_{ab}$ (where $\G = 0$). In particular, there is a
deficit angle $2\pi(1 - \e\! -GH_o/2)$ at infinity.
\cite{9}. As a consequence, {\it we no longer have Poincar\'e
invariance even at infinity} \cite{9,10,11}. There is a preferred
rest-frame given by $\partial/\partial T$ selected by the total
energy-momentum of the system. Perhaps the most dramatic difference
from the Minkowskian situation is that the total Hamiltonian of the
system (including gravity) is now bounded above \cite{10,11}: it is
given by
\be \label{9}
H = \frac{1}{4G}( 1 - e^{-4GH_o})\, .
\ee
In the weak field limit (i.e. as $GH_o(\tilde{C}) \rightarrow 0$), $H$
does tend to $H_o$. However, in the strong field regime, the general
relativistic corrections dominate and the total Hamiltonian is a
bounded, {\it non-polynomial} functional of $H_o$.
 
Let us now switch on both $G$ and $\hbar$, i.e., consider quantum
gravity proper. (Now, we have a natural mass scale ($G^{-1}$) as well
as a natural length scale ($G\hbar$).) As one might expect, the model
can be quantized exactly \cite{6,7,8}. In a Hamiltonian framework, one
can first quantize the field $\Phi$ exactly as in Minkowski space and
then define the metric operator by suitably regularizing the right
side of (\ref{8}). As in any non-trivial model which is solved by
mapping it to a trivial model, the non-trivialities lie in the {\it
relation} between the two models. In the present case, the physical
model leads us to consider, in particular, operators corresponding to
$\e G\G (R,T)$ and study the resulting quantum geometry which exhibits
several interesting features \cite{7}.
  
For simplicity, here we will focus our attention on the asymptotic
form of the metric (whose only non-trivial component is $g_{RR} = -
g_{TT} = \e GH_o$) and on the Hamiltonian $H$. The corresponding
operators will be taken to be:
\be \label{10}
\hat{g}_{RR} = e^{G \hat{H}_o} \quad{\rm and}\quad 
\hat{H} = \frac{1}{4G}(1 - e^{-4G \hat{H}_o}) \, ,
\ee
which are both positive definite and self-adjoint.  (The other
possibility is to further normal order these operators. However, then
the spectrum of the resulting Hamiltonian is unbounded below as well
as above and resemblance with (\ref{9}) is lost. Furthermore, our main
results on quantum fluctuations would not be affected by this change.)

The question now is: Are there semi-classical states which approximate
solutions $(\tilde{C}(R,T), g_{ab})$ of Einstein-Maxwell equations
throughout the space-time? Since we wish to approximate, in
particular, the solution $\tilde{C}$ to the wave equation, the obvious
candidates are, again, the coherent states $\kp$. Let us compute the
expectation values of operators of interest and see if they give us
back the classical fields. Fortunately, these expectation values as
well as fluctuations around them can be expressed in closed forms. We
have:
\ba \label{11}
\bp \hat{\Phi}(R,T) \kp &=& \tilde{C}(R,T);\quad 
\bp \hat{g}_{RR} \kp \, = \, e^{{\frac{1}{\hbar}}} \int d\o\, 
|C(\o )|^2 (e^{G\hbar\o} -1); \nonumber\\
\bp \hat{H}\kp &=& {\frac{1}{4G}}[1 - e^{{\frac{1}{\hbar}} 
\int d\o\, |C(\o )|^2 (e^{-4G\hbar\o} -1)}]\,.
\ea 
Thus, for the scalar field, of course, we recover the classical field.
However, for the metric and the Hamiltonian, the expectation values
are not simply related to the corresponding classical expressions:
\be \label{12}
g_{RR} = e^{G\int d\o\, \o |C(\o )|^2} \quad {\rm and} \quad
H = \frac{1}{4G}\big(1 - e^{-4G\int d\o\, \o |C(\o )|^2} \big)\, .
\ee
(In particular, the expectation values have an explicit $\hbar$
dependence.)

However, since we now have a natural length scale, we can analyze
limits. In the low frequency limit, $G\hbar\o <\!<1$, the
$\hbar$ dependence disappears in the leading order and we recover the
classical values {\it provided the frequency is so small that} $\N
(G\hbar \o)^2 <\!< 1$. In this case, we can also evaluate the quantum
corrections systematically to any desired order:
\ba \label{13}
\bp \hat{g}_{RR} \kp &=& g_{RR}\,\, [1 + \frac{1}{2}\N\, 
(G\hbar \o_o )^2 + ...\,\, ]\, \nonumber\\
\bp \hat{H} \kp &=&  H(\tilde{C}) - \frac{4}{G}\N (G\hbar \o_o)^2 
e^{-4 G H_o (\tilde{C})} + ... \, ,
\ea
where in the last terms, we have evaluated the integrals approximately
for the case when $C(\o)$ is sharply peaked at a frequency $\o_o$.  In
the high frequency regime, $G\hbar\o >\!> 1$, there is always a
significant disagreement with the classical expressions:
\ba \label{14}
\bp \hat{g}_{RR} \kp &=& e^{{\frac{1}{\hbar}}\int d\o\, |C(\o )|^2
e^{G\hbar\o}}\,\,
\approx \,\, e^{\N\, (e^{G\hbar \o_o})}
\nonumber\\
\bp \hat{H} \kp &=&  \frac{1}{4G}[ 1- e^{-{\frac{1}{\hbar}}
\int d\o |C(\o )|^2}] \,\, 
\approx \,\, \frac{1}{4G}[ 1- e^{-\N}]\, ,
\ea
Note that now $\hbar$ does not disappear even in the leading order
terms and the departures from the classical values occur even in the
{\it asymptotic} metric.

Let us analyze the quantum uncertainties. For brevity, we will now
focus just on the metric operator. The exact expression turns out to
be:
\be \label{15}
(\frac{\triangle\hat{g}_{RR}}{<\!\hat{g}_{RR}\!>})^2
=  [e^{{\frac{1}{\hbar}}\int d\o\, |C(\omega)|^2 (1 - e^{G\hbar\o})^2}
\,\, - 1]\, .
\ee
In the low frequency regime, $G\hbar\o <\!<1$, we therefore have:
\be \label{16}
(\frac{\triangle \hat{g}_{RR}}{<\!\hat{g}_{RR}\!>})^2 \approx
e^{\N (G\hbar\o_o )^2}\,\, - 1\ ,
\ee
and, in the high frequency regime, $G\hbar\o >\!>1$, we have:
\be \label{17}
(\frac{\triangle\hat{g}_{RR}}{<\!\hat{g}_{RR}\!>})^2 \approx
e^{\N\, (e^{2G\hbar\o_o})}\, .
\ee
Thus, in the high frequency regime, the quantum fluctuations in the
metric are huge even when the state is sharply peaked at the given
classical scalar field $\tilde{C}$ and they become worse as the number
$\N$ of expected photons increases. (Recall that the opposite holds
for the $\hat{\Phi}$ field.) Furthermore, even if we have a single
photon of Planck frequency  --i.e., a little ``blip'' in $C(\o)$
peaked at the Planck frequency with $\N =1$--  Eq (\ref{15})
implies that $(\triangle\hat g/ <\hat{g}>)^2 > 10^3$! Next, consider
the low frequency regime. Now, for the quantum uncertainties in the
Maxwell field to be negligible, we need $\N >\!>1$ {\it and} for the
uncertainties in the metric to be negligible, we need the frequency to
be so low that $\N (G\hbar \o )^2 <\!<1$. Only in this regime does the
geometry given by classical equations approximate the predictions of
the quantum theory.

To summarize, the quantum theory does have infinitely many states with
semi-classical behavior. This by itself is a non-trivial property,
given the non-linearities of the Einstein equations. However, all
these states are peaked at a very restricted class of classical
solutions: the corresponding Maxwell field profiles $C(\o)$ have to
satisfy the two inequalities: $\N >\!>1$ and $\N (G\hbar\o)^2
<\!<1$. Thus, most classical solutions  --including the one
representing the gravitational field caused by a ``blip'' of
(trans-)Planckian frequency in the Maxwell field--  do not arise as
classical limits of the quantum theory; they fail to serve as
leading order approximations even in regions where the curvature is
small. This is a subtle failure of the classical theory;
non-linearities of Einstein's equations magnify the small quantum
uncertainties in the Maxwell field to huge fluctuations in the metric.
It would have been difficult to guess this process before hand and
arrive at the two specific restrictions from general considerations.

We will conclude by summarizing some ramifications of these results.

1. {\it Semi-classical gravity:} Recall \cite{4} that a solution to
semi-classical gravity consists of a quadruplet $(M, g_{ab}, \hat\Phi,
|\Psi\!>)$ consisting of a metric $g_{ab}$ (of hyperbolic signature)
on a manifold $M$, a quantum matter field $\hat{\Phi}$ satisfying a
field equation on $(M, g_{ab})$ and a state $|\Psi\!>$ of the field
such that the semi-classical Einstein equation $G_{ab} = 8\pi G
<\!\Psi| \hat{T}_{ab}|\Psi\!>$ is satisfied. In 4-dimensions, it is
difficult to find exact solutions to this theory and several important
issues about the nature of the approximation involved remain open.
What is the situation in the present model? Now, it is straightforward
to show that the theory admits an infinite number of
solutions. Indeed, given {\it any} (axi-symmetric) classical solution
$(\tilde{C}, g_{ab})$ to Einstein-Maxwell equations on $M= R^3$, the
quadruplet $(M, g_{ab}, \hat{\Phi}, \kp)$ is an exact solution to
semi-classical gravity. While this abundance serves to show that the
theory is mathematically rich, it also brings out its physical
limitations.  For, unless the profiles $C(\o)$ of the Maxwell field
satisfy $\N (G\hbar \o )^2 <\!<1$, the solution to semi-classical
gravity is spurious; it does not approximate the situation in the full
theory. Very roughly, it knows about both $G$ and $\hbar$ separately,
but does not exploit the Planck length which requires {\it both} at
once.

2. {\it High frequency behavior:} We saw that something peculiar does
happen at Planckian and trans-Planckian frequencies. However, {\it
local} Lorentz invariance is not broken; the field equation governing
$\tilde{C}$ does not, for example, involve an additional vector field.
The frequency refers to the asymptotic rest-frame ($\partial/\partial
T$).  The relativistic dispersion relation is not modified, nor is
there a high frequency cut-off. Similarly, within this model, there is
no hint of the behavior suggested by the holographic hypothesis.  It
is sometimes conjectured that in a full quantum theory the
trans-Planckian strangeness will trickle down and even affect the
vacuum since the vacuum fluctuations involve all modes. There is no
evidence of such a behavior either. The vacuum $|0\!>$ is stable
although the vacuum fluctuations involve modes of arbitrarily high
frequencies. Furthermore, the vacuum is an eigenstate both of
$\hat{H}$ (with eigenvalue zero) and of $\hat{g}_{RR} = -\hat{g}_{TT}$
(with eigenvalue one).  The peculiarity that does arise is of a rather
different nature: {\it classical geometry simply fails to be a good
approximation} when $C(\o)$ has a significant high frequency component
and one has to take the quantum nature of geometry  --with all its
fluctuations--  seriously.

3. {\it 3 versus 4 dimensions:} All our discussion is based on a
specific model. What are its main limitations?  The model refers to
3-dimensional gravity which has several peculiarities. First, since
Newton's constant in three dimensions has physical dimensions of
inverse mass (rather than length times inverse mass) we can not
associate a Schwarzschild radius to a given mass (without introducing
a cosmological constant). This explains intuitively why we did not
encounter gravitational collapse. If we consider a spherical scalar
field in 4 dimensions, on the other hand, the theory has two sectors;
one leading to black holes (strong data) and the other leading to
simply scattering (weak data). The present model can only give us
insights in to the second sector. Hence, it would be imprudent to draw
from it definitive conclusions about the Bekenstein bound \cite{2} and
related conjectures \cite{3}.  A second difference is that (in the
asymptotically flat context) while in 3 dimensions the metric outside
sources is flat, in 4-dimensional general relativity it falls off as
$1/r$. Therefore, in 4-dimensional asymptotically flat situations, the
effect of trans-Planckian frequencies will decay and the fluctuations
will be significant only near (but possibly in a macroscopic region
around) the sources. In spite of these limitations, the model does
provide concrete evidence for large quantum gravity effects in domains
that have been generally ignored. Furthermore, they are not artifacts
of an unusual quantization procedure; we used only traditional
Fock-space methods.  Therefore, with appropriate modifications,
features encountered here should persist in more sophisticated
models. Indeed, as already remarked, the model arises directly from
symmetry reduction of {\it 4-dimensional} general relativity \cite{5}
and similar reductions exist also in string theory \cite{13}.

Details as well as several other issues will be discussed in
\cite{12}.

\bigskip
\bigskip
{\bf Acknowledgements:} I would like to thank Chris Beetle, Eanna
Flanagan, Donald Marolf, Monica Pierri and Thomas Thiemann for
discussions. This work was supported in part by the NSF grant
95-14240, by the Eberly research funds of Penn State University and by
the Erwin Schr\"odinger International Institute for Mathematical
Sciences, Vienna.

\end{document}